\documentclass[12pt]{article}  
\makeatletter
\def\fmslash{\@ifnextchar[{\fmsl@sh}{\fmsl@sh[0mu]}}
\def\fmsl@sh[#1]#2{%
  \mathchoice
    {\@fmsl@sh\displaystyle{#1}{#2}}%
    {\@fmsl@sh\textstyle{#1}{#2}}%
    {\@fmsl@sh\scriptstyle{#1}{#2}}%
    {\@fmsl@sh\scriptscriptstyle{#1}{#2}}}
\def\@fmsl@sh#1#2#3{\m@th\ooalign{$\hfil#1\mkern#2/\hfil$\crcr$#1#3$}}
\makeatother
\global\arraycolsep=2pt 
\input{epsf}
\usepackage{bbm}
\usepackage{a4wide}
\usepackage{latexsym}
\begin{document} 
\thispagestyle{empty}
\rightline{TTP02-29}
\rightline{hep-ph/0210248}
\rightline{\today}
\bigskip
\boldmath
\begin{center}
{\bf \Large 
Subleading Light-cone distributions functions \\[3mm] in the decays
$\Lambda_b \to X_s \gamma$ and $\Lambda_b \to X_u \ell \bar{\nu}_\ell$}
\end{center}
\unboldmath
\smallskip
\begin{center}
{\large{\sc Michael Kraetz} and {\sc Thomas Mannel}}
\vspace*{2cm} \\ 
{\sl Institut f\"{u}r Theoretische Teilchenphysik, \\
Universit\"{a}t Karlsruhe,  D--76128 Karlsruhe, Germany}
\end{center}
\begin{abstract}
\noindent
In this paper we investigate the subleading twist corrections
to the photon energy spectrum in the decay
$\Lambda_b \to X_s \gamma$ and to the lepton energy spectrum
in $\Lambda_b \to X_u \ell \bar{\nu}_\ell$.
As a first step we rederive the matching coefficients
of the subleading twist using a much simpler method. Parametrizing the
matrix elements for $\Lambda_b$ and using reparametrization
invariance, we show that the energy spectra close to the endpoints
are given in terms of only a single universal function.
\end{abstract}
\newpage
\section{Introduction}
The existence of baryons with bottom quarks is well established by now,
and at future facilities a detailed study of the decay
modes of such states will become feasible. However, the $B$ factories
are running below the threshold for the production of these modes, which
means that the measurements we are going to discuss will be -- at least
in the foreseeable future -- the domain of hadron colliders. 

From the theoretical point of
view, the ground state bottom baryon $\Lambda_b$ is the simplest heavy
state, although it contains three quarks. The reason is that the spin of the
heavy $b$ quark decouples in the limit $m_b \to \infty$ and thus the light
degrees of freedom have to be in a spin-zero state \cite{Baryons}.
In other words, the
spin of a $\Lambda_b$ is the spin of the heavy quark, which in principle
allows interesting polarization studies. Some of these studies require
polarized $\Lambda_b$'s which may become available at GIGA-Z \cite{TeslaTDR},
a high-luminosity linear collider running at the $Z_0$ resonance.

In the present paper we investigate energy spectra of
inclusive heavy-to-light decays of the $\Lambda_b$, in particular the 
photon spectrum of the rare decay $\Lambda_b \to X_s \gamma$ and 
the lepton energy spectrum of the semileptonic decay 
$\Lambda_b \to X_u \ell \bar{\nu}_\ell$. 
The situation here is very similar to the one in the decays 
$B \to X_s \gamma$ and $B \to X_u \ell \bar{\nu}_\ell$: 
The photon spectrum in $B \to X_s \gamma$ is mainly concentrated in the
endpoint of small hadronic mass of $X_s$, where the strict $1/m$ expansion
breaks down. This forces us to switch to an expansion in twists similar
to deep inelastic scattering. The leading twist contribution is given
in terms of a light-cone-distribution function which can be
formally written as a forward matrix element of a non-local
operator \cite{shape}.
This function describes all inclusive heavy to light transitions and 
is universal. Applying this to $b \to u \ell \bar{\nu}_\ell$
and $b \to s \gamma$ one may obtain a model independent determination 
of $|V_{ub} / V_{ts}|$ \cite{VubNeubert}.

Recently subleading terms have been included into the analysis
of $B$ meson decays \cite{BLM1}, and the purpose of the present paper
is to include subleading terms for the inclusive decays of heavy baryons.
The main
result of the paper is that only a single distribution function is
needed to parametrize the differential decay rates to subleading order;
in other words, the endpoints of the photon energy spectrum and the
charged lepton energy spectrum in $\Lambda_b \to X_s \gamma$ and
$\Lambda_b \to X_u \ell \bar{\nu}_\ell$ are given in terms of one
universal function up to terms of order $1/m_b^2$. This has to be compared
to the case of the corresponding $B$ decays, where three universal functions
are needed \cite{BLM1}. 

In the next section we consider the matching including the terms of
subleading twist. We reproduce the results of \cite{BLM1} using a much
simpler method based on a background field, in which the light quark
propagates. In section 3 we parametrize the forward
matrix elements for the non-local operators between $\Lambda_b$ states. 
In section 4 we discuss our result and conclude.

\section{Matching on non-local operators to order $1/m$}
The tree-level matching of full QCD to the non-local operators
to order $1/m_b$ has been performed for both $b\to s \gamma$ \cite{BLM1}
and $b \to u \ell \bar{\nu}_\ell$ \cite{BLM2, LLW1} and the results of these
papers can  be used also for the case of the $\Lambda_b$. However, we find
it useful to rederive the results of $\cite{BLM1}$ in a different way,
which is very transparent and is manifestly gauge invariant. 
The method resembles the one used in \cite{ManNeu} for the shape function
for mesons. 

The starting point is the usual decomposition of the heavy quark momentum
according to
\begin{equation} 
p_b = m_b v + k 
\end{equation}
where the components of the residual momentum $k$ are small compared to
$m_b$. At tree level, the inclusive decay $b \to s \gamma$ requires
to consider the propagator of the (massless) $s$ quark, which is 
\begin{equation}
S = \frac{i}{\fmslash{p}_b - \fmslash{q}} =
    \frac{i}{m_b \fmslash{v} + \fmslash{k} - \fmslash{q}}
\end{equation}
where $q$ is the photon momentum 

In position space and in terms of the static heavy-quark field
$h_v (x)$ the residual momentum $k$ corresponds to a covariant
derivative, since the large part $m_b v$ has been removed by a field
redefinition. This leads us to consider the following
(non-local) operator
\begin{equation} \label{first}
T =  \bar{h}_v (0) \Gamma^\dagger \left(
     \frac{i}{m_b \fmslash{v} + i \fmslash{D} - \fmslash{q}} \right)
     \Gamma h_v (x) 
\end{equation}
where $D$ is the (QCD) covariant derivative and $\Gamma$ is the Dirac matrix
of the corresponding heavy-to-light current $\bar{s} \Gamma b$ 

It has been shown in \cite{ManNeu} that (\ref{first}) yields the correct
tree level matching for the shape function and we shall use the same
arguments to rederive the subleading terms. In order to define the power
counting properly we introduce light-cone vectors as
\begin{equation}
v = \frac{1}{2} (n + \bar{n}) , \quad  n^2 = 0 = \bar{n}^2 , \quad
n \cdot \bar{n} = 2 , \quad  q = (v \cdot q) \bar{n}  
\end{equation}
In this way we can decompose any vector into its components according to
\begin{equation}
g_{\mu\nu}= \frac{1}{2} n_\mu \bar{n}_\nu+ \frac{1}{2}\bar{n}_\mu n_\nu
+g_{\mu\nu}^\perp 
\end{equation}
We want to expand (\ref{first}) in the kinematical region where
\begin{equation}
(m_b v - q) \cdot n = m - n \cdot q \sim \Lambda_{QCD} 
\end{equation}
which is the kinematical region for the shape function.
In order to perform this expansion we write 
\begin{equation}
\frac{1}{m_b \fmslash{v} + i \fmslash{D} - \fmslash{q}} =
\frac{m_b \fmslash{v} + i \fmslash{D} - \fmslash{q}}
    {(m_b \fmslash{v} + i \fmslash{D} - \fmslash{q})^2}
\end{equation}
and consider the denominator, which can be written in the form
\begin{equation}
(m_b \fmslash{v} + i \fmslash{D} - \fmslash{q})^2 =
m_b Q_+ + \frac{1}{2} \{iD_-,Q_+\}  + (iD^\perp)^2 -
i\sigma^{\mu\nu} iD_\mu iD_\nu 
\end{equation}
where we have defined 
\begin{equation}
Q_+ = m_b  - n \cdot q + i D_+  , \quad i D_+ = i (n \cdot D) , \quad
i D_- = i (\bar{n} \cdot D), \quad D_\mu^\perp = g_{\mu \nu}^\perp D^\nu
\end{equation}
Since $m_b - n \cdot q \sim \Lambda_{QCD}$ we have $Q_+ \sim \Lambda_{QCD}$
and thus the term $m_B Q_+$ is the leading term (of order $m_b \Lambda_{QCD}$)
in the denominator. All other terms are of order $\Lambda_{QCD}^2$ and
are expanded as 
\begin{eqnarray}
\frac{1}{(m\fmslash{v}-\fmslash{q}+i\fmslash{D})^2} & = & \frac{1}{mQ_+} -
\frac{1}{mQ_+}\frac{1}{2}\{iD_-,Q_+\}\frac{1}{mQ_+} \\\nonumber
&& -\frac{1}{mQ_+}((iD^\perp)^2 - i\sigma^{\mu\nu}iD_\mu
iD_\nu)\frac{1}{mQ_+} + \ldots
\end{eqnarray}
Next we reinsert the numerator, which we do in a symmetrized fashion. 
After some simple algebra we get 
\begin{eqnarray}
\frac{m\fmslash{v}-\fmslash{q}+i\fmslash{D}}
{(m\fmslash{v}-\fmslash{q}+i\fmslash{D})^2}
&=& \frac{\fmslash{n}}{2Q_+} + \frac{\fmslash{\bar{n}}}{2m}+
\frac{1}{2m}\left\{i\fmslash{D}^\perp,\frac{1}{Q_+}\right\} \\
&& - \frac{1}{2m}\frac{1}{Q_+}\left(\fmslash{n}(iD^\perp)^2
   -\frac{1}{2}\{\fmslash{n},i\sigma^{\mu\nu}\}iD_\mu
iD_\nu\right)\frac{1}{Q_+} \nonumber
\end{eqnarray}
The last term can be rewritten in terms of the perpendicular components of the
covariant derivative. We have 
\begin{eqnarray}
\{\fmslash{n},\sigma^{\mu \nu}\} n_\mu n_\nu &=& \frac{i}{2}
(\fmslash{n}(\fmslash{n}\fmslash{\bar{n}}-\fmslash{\bar{n}}\fmslash{n})
+
(\fmslash{n}\fmslash{\bar{n}}-\fmslash{\bar{n}}\fmslash{n})\fmslash{n})
= 0  \\
\{\fmslash{n},\sigma^{\mu \nu}\} n_\mu g_{\nu \alpha}^\perp &=& \frac{i}{2}
(\fmslash{n}(\gamma^\perp_\alpha\fmslash{n}-\fmslash{n}\gamma^\perp_\alpha) = 0
\end{eqnarray}
from which we derive
\begin{equation}
\{\fmslash{n},i\sigma^{\mu\nu}\}iD_\mu iD_\nu =
i\sigma^{\alpha \beta} n_\alpha \bar{n}_\beta
\gamma_\perp^\mu[iD^\perp_\mu,Q_+] +
2\fmslash{n}i\sigma^{\mu\nu}_\perp iD_\mu^\perp iD_\nu^\perp
\end{equation}
Combining this with the other terms we get
\begin{eqnarray} \label{final1}
\frac{m \fmslash{v}-\fmslash{q}+i\fmslash{D}}
     {(m\fmslash{v}-\fmslash{q}+i\fmslash{D})^2} &=&
\frac{\fmslash{n}}{2Q_+} + \frac{\fmslash{\bar{n}}}{2m} 
+ \frac{\gamma^\mu_\perp}{2m}\left\{iD_\mu,\frac{1}{Q_+}\right\}
+\frac{i\sigma^{\alpha \beta} n_\alpha \bar{n}_\beta
\gamma_\perp^\mu}{4m}\left[iD_\mu,\frac{1}{Q_+}\right] \nonumber \\
& - &
\frac{1}{2m}\frac{\fmslash{n}}{Q_+}\left(\frac{1}{2}g^{\mu\nu}_\perp\{iD_\mu^\perp,iD_\nu^\perp\} 
-\frac{1}{2}i\sigma^{\mu\nu}_\perp [iD_\mu^\perp,iD_\nu^\perp]\right)
\frac{1}{Q_+} 
\end{eqnarray}

This result can now be used to perform the matching on the non-local
operators as they have been defined in \cite{BLM1}. These operators are
\begin{eqnarray}
O_0 (\omega) &=& \bar{h}_v \delta(\omega+i\hat{D}_+) h_v \\
O_1^\mu(\omega) &=& \bar{h}_v\left\{iD_\mu,\delta(\omega+i\hat{D}_+)\right\}
h_v \\
O_2^\mu(\omega) &=& \bar{h}_v\left[iD_\mu,\delta(\omega+i\hat{D}_+)\right]
h_v \\
O_3^{\mu\nu}(\omega_1,\omega_2) &=& \bar{h}_v
\delta(\omega_2+i\hat{D}_+)\{iD^\mu_\perp,iD^\nu_\perp\}
\delta(\omega_1+i\hat{D}_+) h_v \\
O_4^{\mu\nu}(\omega_1,\omega_2) &=& \bar{h}_v
\delta(\omega_2+i\hat{D}_+)[iD^\mu_\perp,iD^\nu_\perp]
\delta(\omega_1+i\hat{D}_+) h_v
\end{eqnarray}
and
\begin{eqnarray}
P_{0,\alpha}(\omega) &=& \bar{h}_v
\delta(w+i\hat{D}_+)\gamma_\alpha\gamma_5 h_v \\
P_{1,\alpha}^\mu(\omega) &=&
\bar{h}_v\left\{iD_\mu,\delta(\omega+i\hat{D}_+)\right\}
\gamma_\alpha \gamma_5 h_v \\
P_{2,\alpha}^\mu(\omega) &=&
\bar{h}_v\left[iD_\mu,\delta(\omega+i\hat{D}_+)\right]
\gamma_\alpha \gamma_5 h_v \\
P_{3,\alpha}^{\mu\nu}(\omega_1,\omega_2) &=& \bar{h}_v
\delta(\omega_2+i\hat{D}_+)\{iD^\mu_\perp,iD^\nu_\perp\}
\delta(\omega_1+i\hat{D}_+)\gamma_\alpha \gamma_5 h_v \\
P_{4,\alpha}^{\mu\nu}(\omega_1,\omega_2) &=& \bar{h}_v
\delta(\omega_2+i\hat{D}_+)[iD^\mu_\perp,iD^\nu_\perp]
\delta(\omega_1+i\hat{D}_+)\gamma_\alpha \gamma_5 h_v
\end{eqnarray}

The matching onto these operators is performed for the decay rate. 
In order to obtain the rate we have to take the imaginary part of
a forward matrix element with the $\Lambda_b$. Inserting (\ref{final1})
back into (\ref{first}) and taking the imaginary part we obtain for the
leading term 
\begin{eqnarray}
{\rm Im}\, \bar{h}_v
\frac{\bar{\Gamma}\fmslash{n}\Gamma}{2Q_+} h_v
&=& -\frac{1}{2m} \int d\omega \, \left(\frac{\pi}{2}
Tr\left[P_+\Gamma\fmslash{n}\Gamma\right]\delta(1-x-\omega) \bar{h}_v
\delta(w+i\hat{D}_+) h_v \right. \nonumber \\
&& \left.- \frac{\pi}{2} Tr\left[s^\alpha\Gamma\fmslash{n}\Gamma\right]
\delta(1-x-\omega) \bar{h}_v
\delta(w+i\hat{D}_+)\gamma_\alpha\gamma_5 h_v \right) \nonumber \\
&=& -\frac{1}{2m} \int d\omega \,
[C_0(\omega)O_0(\omega)+C_{5,0}^\alpha(\omega)P_{0,\alpha}(\omega)]
\end{eqnarray}
from which we can read off the matching coefficients $C_0$ and
$C_{5,0}^\alpha$ as they have been derived already in
\cite{BLM1}\footnote{Here we made us of the decomposition formula
$$ P_+\Gamma P_+ = \frac{1}{2}P_+ Tr[P_+\Gamma] - \frac{1}{2}s^\mu
Tr[s_\mu\Gamma] \qquad s_\mu = P_+ \gamma_\mu \gamma_5 P_+ \qquad P_+=\frac{1}{2}(1+\fmslash{v}). $$}

In the same way we obtain for the subleading contributions
\begin{eqnarray}
&& {\rm Im} \, \bar{h}_v \frac{\bar{\Gamma}\gamma^\mu_\perp\Gamma}{2m}
         \left\{iD_\mu,\frac{1}{Q_+}\right\} h_v \\ \nonumber
&=&  {\rm Im} \int d\omega \,
\bar{h}_v \frac{\bar{\Gamma}\gamma^\mu_\perp\Gamma}{2m^2}
\{iD_\mu,\delta(\omega+i\hat{D}_+)\}\frac{1}{1-x-\omega} h_v \\ \nonumber
&=& -\frac{1}{2m^2} \int d\omega \,
\left(\frac{\pi}{2}Tr\left[P_+\bar{\Gamma}\gamma^\mu_\perp\Gamma\right]
      \nonumber \delta(1-x-\omega)
\bar{h}_v\{iD_\mu,\delta(\omega+i\hat{D}_+)\} h_v \right.
\\ \nonumber
&& \qquad \left.
-\frac{\pi}{2}Tr\left[s^\alpha\bar{\Gamma}\gamma^\mu_\perp\Gamma\right]
\delta(1-x-\omega)
\bar{h}_v \{iD_\mu,\delta(\omega+i\hat{D}_+)\}\gamma_\alpha\gamma_5 h_v \right)
\end{eqnarray}
from which we obtain the matching coefficients $C_1^\mu$ and
$C_{5,1}^{\alpha,\mu}$

A further restriction of the structure of the matching coefficients, which should hold even beyond tree level, is due to reparametrization
invariance \cite{LukeManohar}. It turns out that reparametrization
invariance \cite{CM,Manohar} relates the coefficients of $ O_0$ and
$O_3^{\mu \nu}$ in such a way that only one function is needed
to parametrize the effect of the two operators $O_0$ and
$O_3^{\mu \nu}$. We shall use the relations from \cite{CM} in the
next section when we will count the number of of unknown functions
needed to describe the spectrum of $\Lambda_b \to X_s \gamma$ including
subleading contributions.

In addition to these operators  we have also to take into account 
the subleading terms of the Lagrangian. They are given as
\begin{eqnarray}
O_T (\omega) &=& i \int d^4x \int \frac{dt}{2 \pi} e^{-i\omega t}
\, {\rm T}[ \bar{h}_v(0)  h_v (t) {\cal L}_{1/m}(x) ] \\
P_{T,\alpha}  (\omega)
&=& i \int d^4x \int \frac{dt}{2 \pi} e^{-i\omega t}
\, {\rm T}[ \bar{h}_v(0) \gamma_\alpha \gamma_5  h_v (t) {\cal L}_{1/m} (x) ] 
\end{eqnarray}
where ${\cal L}_{1/m}$ is the subleading contribution to the Lagrangian
\begin{equation}
{\cal L}_{1/m} =  \frac{1}{2 m} \bar{h}_v (i \fmslash{D})^2 h_v 
\end{equation}

Once matrix elements are taken, these pieces can be interpreted as the
corrections to the states, which means that they will appear always in a
specific combination with the leading terms, i.e. these will not introduce
any new unknown functions.

\section{Matrix Elements for $\Lambda_b$ Decays}
After having computed the matching we have to consider the forward
matrix elements with a $\Lambda_b$. These matrix elements have to be
evaluated in the static limit in which the $\Lambda_b$ becomes a very
simple object. We obtain for the matrix elements:
\begin{eqnarray}
\langle \Lambda_b (v,s) | O_0(\omega) |  \Lambda_b (v,s) \rangle
&=& \bar{u} (v,s) u(v,s)  f_\Lambda (\omega) \\
\langle \Lambda_b (v,s) | P_{0,\alpha} (\omega) |  \Lambda_b (v,s) \rangle
&=& \bar{u} (v,s) \gamma_\alpha \gamma_5 u(v,s)  f_\Lambda (\omega) 
\end{eqnarray}
where $u(v,s)$ describes the static $\Lambda_b$ with velocity
$v$ and spin $s$.

For the subleading operators the non-vanishing
matrix elements are parametrized as
\begin{eqnarray}
&& \langle \Lambda_b (v,s) | O_1^\mu (\omega) |  \Lambda_b (v,s) \rangle
= -2 \bar{u} (v,s) u(v,s) (v^\mu - n^\mu) \omega  f_\Lambda (\omega) \\
&& \langle \Lambda_b (v,s) |
P_{1,\alpha}^\mu (\omega) |  \Lambda_b (v,s) \rangle
= -2\bar{u} (v,s) \gamma_\alpha \gamma_5 u(v,s)
    (v^\mu - n^\mu) \omega  f_\Lambda (\omega) \\
&& \langle \Lambda_b (v,s) | O_3^{\mu \nu} (\omega_1, \omega_2)
                        |  \Lambda_b (v,s) \rangle
= \bar{u} (v,s) u(v,s) g_\perp^{\mu \nu} g_\Lambda (\omega_1, \omega_2) \\
&& \langle \Lambda_b (v,s) | P_{3,\alpha}^{\mu \nu} (\omega_1, \omega_2)
                        |  \Lambda_b (v,s) \rangle 
= \bar{u} (v,s) \gamma_\alpha \gamma_5 u(v,s)
    g_\perp^{\mu \nu} g_\Lambda (\omega_1, \omega_2) \\
&& \langle \Lambda_b (v,s) | O_T (\omega)|  \Lambda_b (v,s) \rangle
= \bar{u} (v,s) u(v,s) t_\Lambda (\omega) \\ 
&& \langle \Lambda_b (v,s) | P_{T,\alpha} (\omega) | \Lambda_b (v,s) \rangle 
= \bar{u} (v,s) \gamma_\alpha \gamma_5 u(v,s) t_\Lambda (\omega) 
\end{eqnarray}

Consequently, only two additional functions appear at order $1/m_b$.
Furthermore, these subleading function are tied to the leading order
function  $f_\Lambda$ by reparametrization invariance, which means that
it appears always the combination
\begin{equation}
F_\Lambda(\omega) = f_\Lambda(\omega) + t_\Lambda (\omega) - \frac{1}{m_b^2} 
\int\! d\omega_1  \, d\omega_1 \, g_\Lambda (\omega_1 \omega_2)  
\left(\frac{\delta(\omega - \omega_1) -
            \delta(\omega - \omega_2)}
            {\omega_1 - \omega_2} \right)
\end{equation}
with the leading order function $f_\Lambda$. In other words, due to
heavy quark spin symmetry we can express $\Lambda_b \to X_s \gamma$
and $\Lambda_b \to X_u \ell \bar{\nu}_\ell$ in terms of a single
universal function, up to terms of order $1/m_b^2$.  

Finally we discuss the moment expansion of the resulting functions.
For the leading term we have
\begin{equation}
f_\Lambda(\omega) = \delta (\omega)
- \frac{\lambda_\Lambda}{6 m_b^2} \delta^{\prime \prime} (\omega) + \cdots
\end{equation}
where
\begin{equation}
\lambda_\Lambda \, \bar{u} (v,s) u(v,s) =
\langle \Lambda_b (v,s) | \bar{h}_v (iD)^2 h_v | \Lambda_b (v,s) \rangle 
\end{equation}
is the kinetic energy of the $b$ quark inside the $\Lambda_b$ baryon.
For the subleading contributions we obtain
\begin{eqnarray}
g_\Lambda (\omega_1, \omega_2) &=& \frac{2 \lambda_\Lambda}{3}
                                \delta (\omega_1)  \delta (\omega_2) + \cdots
\\
t_\Lambda (\omega) &=& - \frac{\lambda_\Lambda}{2m_b^2} \delta^\prime (\omega)
                   + \cdots
\end{eqnarray}
which reproduces the known result from \cite{BauerMom} with the replacements
$\lambda_1 \to \lambda_\Lambda$ and $\lambda_2 \to 0$:
\begin{equation}
\frac{d \Gamma}{dx} (\Lambda_b \to X_s \gamma) =
\Gamma_0 \left[\delta(1-x)
         - \frac{\lambda_\Lambda}{2 m_b^2} \delta^\prime (1-x)
         - \frac{\lambda_\Lambda}{6 m_b^2} \delta^{\prime \prime} (1-x)
         + \cdots \right]
\end{equation}
with
\begin{equation}
\Gamma_0 = \frac{G_F^2 \alpha |V_{tb}^* V_{ts}| |C_7|^2}
             {32 \pi^4} m_b^5 
\end{equation}
The moment expansion of the new universal function reads
\begin{equation} \label{FMoments}
F_\Lambda (\omega) = \delta(\omega) 
         + \frac{\lambda_\Lambda}{6m_b^2} \delta^\prime (\omega) + \cdots
\end{equation}

\section{Discussion and Conclusions}
The main result of this paper is that the semileptonic
and radiative decays of a $\Lambda_b$ baryon into light hadrons
are described by a universal function, where the corrections to
this statement are of order $1/{m_b}^2$. In this way we find
the two differential decay rates:
\begin{eqnarray} \label{Lam2xsgamma}
\frac{d \Gamma}{d x} (\Lambda_b \to X_s \gamma)
      &=& \frac{G_F^2 \alpha |V_{tb}^* V_{ts}| |C_7|^2}
             {32 \pi^4} {m_b}^5 (2x - 1) F_\Lambda (1-x) \\
\label{Lam2Xuell}
\frac{d \Gamma}{d y} (\Lambda_b \to  X_u \ell \bar{\nu}_\ell)
      &=& \frac{G_F^2 |V_{ub}|^2}{96 \pi^3} {m_b}^5
          \int d\omega \, \Theta(1-y-\omega)  (1-\omega)  F_\Lambda (\omega)
\end{eqnarray}
where $x = 2 E_\gamma /m_b$ is the rescaled photon energy and
$y =2 E_\ell /m_b$ is the rescaled lepton energy. Note that the first moment
of $F$ still vanishes up to terms of order $1/{m_b}^2$, so we still get
no contribution to the total rates. This serves a a check, since
the contributions of order $1/m$ may not contribute to the rate. 
As for $\Lambda_b \to X_s \gamma$, the spectrum of the semileptonic
decay can be expanded in terms of moments. Inserting (\ref{FMoments}) into
the spectrum, one arrives at
\begin{equation}
\frac{d\Gamma}{dy}(\Lambda_b \to  X_u \ell \bar{\nu}_\ell)=
\Gamma_0\left[\theta(1-y)-\frac{\lambda_\Lambda}{6{m_b}^2}\delta(1-y)
-\frac{\lambda_\Lambda}{6{m_b}^2}\delta^\prime(1-y)+\ldots\right]
\end{equation}
where
\begin{equation}
\Gamma_0=\frac{G_F^2 |V_{ub}|^2}{96 \pi^3} {m_b}^5.
\end{equation}
With the replacements already mentioned above, this agrees with the
result obtained in \cite{BauerMom}. 

Although this may be difficult from the experimental point of view,
we may investigate
the subleading corrections to a determination of $V_{ub}$ from a
comparison of $\Lambda_b \to X_s \gamma$ with
$\Lambda_b \to X_u \ell \bar{\nu}_\ell$. Like in the case of $B$ mesons,
an energy cut $E_c$ on the charged lepton will be unavoidable to discriminate
the large charm background, i.e.
$E_c > (M^2 [\Lambda_b] - M^2 [\Lambda_c])/(2 M[\Lambda_b])$. One may
define observables (similar as for the decays of $B$ mesons,
see \cite{Rothstein}) involving partially integrated rates with suitable
weight functions. Since the $1/m_b$ terms appear only a kinematic
factors (the pre-factors $(2x-1)$ in (\ref{Lam2xsgamma}) and $(1-\omega)$
in (\ref{Lam2Xuell})) we have more complicated weight functions as in
\cite{Rothstein}. We define 
\begin{eqnarray}\label{GuGsdefs}
\Gamma_u(E_c) &\equiv&
\int_{E_c}^{m_\Lambda/2} d E_\ell \left(\frac{4{E_\ell}^2 - E_c m_b}{2{E_\ell}^2}\right)
\frac{d \Gamma_u^{\Lambda_b \to X_u \ell \bar{\nu}_\ell}}{d E_\ell}
\nonumber\\
\Gamma_s(E_c) &\equiv&
\frac{2}{m_b} \int_{E_c}^{m_\Lambda/2} d E_\gamma (E_\gamma - E_c)
\frac{d \Gamma_s^{\Lambda_b \to X_s \gamma}}{d E_\gamma} \,.
\end{eqnarray}
from which $V_{ub}$ can be determined as 
\begin{equation} \label{Vub}
\left|\frac{V_{ub}}{V_{tb}^{}V_{ts}^*}\right| = 
\frac{3 \alpha}{\pi} |C_7^{\rm eff}|^2
\frac{\Gamma_u(E_c)}{\Gamma_s(E_c)} + {\cal O} (1/{m_b}^2)
\end{equation}

Comparing our result (\ref{Vub}) for the $V_{ub}$ determination
with the one obtained in \cite{BLM2,LLW1} one expects 
substantially smaller corrections in the case of $\Lambda_b$ baryons,
since the corrections are only of order $1/{m_b}^2$.
However, this requires to measure the inclusive semileptonic and
radiative rare $\Lambda_b$ decays, which is more difficult experimentally.

\section*{Acknowledgements}
This work was supported by the DFG Graduiertenkolleg ``Hochenergiephysik and
Teilchen\-astrophysik'', by the DFG Forschergruppe ``Quantenfeldtheorie,
Computeralgebra und Monte Carlo Simulationen'' and by the Ministerium f\"ur
Bildung und Forschung bmb+f.

\end{document}